\documentclass[12pt]{article}
 \usepackage{graphicx}
 \usepackage[cp1251]{inputenc}

\begin{document}
 \noindent {\footnotesize\it
Astronomy Reports, 2022, Vol. 66, No. 7, pp. 545–554.
 \newcommand{\dif}{\textrm{d}}
 }

 \noindent
 \begin{tabular}{llllllllllllllllllllllllllllllllllllllllllllll}
 & & & & & & & & & & & & & & & & & & & & & & & & & & & & & & & & & & & & & &\\\hline\hline
 \end{tabular}

  \vskip 0.5cm
  \bigskip
\centerline{\bf\Large Analysis of the Distance Scales by Cepheids}
 \centerline{\bf\Large from the Gaia EDR3 Catalogue Data}

 \bigskip
 \bigskip
 \centerline{\bf
            V. V. Bobylev\footnote[1]{e-mail: vbobylev@gaoran.ru} (1) and
            A. T. Bajkova (1)
            }
 \bigskip
 \centerline{\small \it(1)
 Pulkovo Astronomical Observatory, Russian Academy of Sciences, St. Petersburg, 196140 Russia}

\bigskip

\centerline{\bf Received March 8, 2022; revised April 3, 2022; accepted May 16, 2022}

 \bigskip
 \bigskip
\noindent {\bf Abstract}--- We study the kinematics of a sample of classical Cepheids younger than 120 Myr. For these stars, the estimates of distances taken from Skowron et al., which are based on the period–luminosity relation, and the line-of-sight velocities and the proper motions from the Gaia catalog are available. There are also distance estimates derived from the trigonometric parallaxes contained in the Gaia ERD3 catalog. A method, which relies on comparison of the first-order derivative of the Galactic rotation angular velocity, showed the need to lengthen the distance scales determined by Skowron et al. by about 10\%. This conclusion was confirmed by direct comparison to the distances predicted on using the trigonometric parallaxes. With taking into account this result, we obtained new estimates of the Galactic rotation parameters and the parameters of a spiral density wave.

\bigskip\noindent
{\bf DOI:} 10.1134/S1063772922080029


\newpage
 \section*{INTRODUCTION}
Cepheids are stars, the brightness variability of
which is caused by their radial pulsations. By means of
the period–luminosity relation [1, 2], it is possible to
estimate the distances to these stars with high accuracy.
Ultimately, this allows one to establish an independent
Cepheid distance scale and to cover with it a
substantial area of the Universe. Classical Cepheids
are those exhibiting pulsation periods of 1 to 100 days.
Their age does not exceed approximately 400 Myr.
These stars are important for studying the properties
of the thin disk of the Galaxy, its spiral structure, rotation,
evolution, etc.

For example, with the use of 220 classical Cepheids
with the proper motion values from the Hipparcos catalog
[3], the Galactic rotation parameters were estimated
with higher precision [4, 5]. Different samples
of classical Cepheids were used to estimate the Galactic
spiral structure parameters [6–10]. Since Cepheids
are seen from large distances, these stars are used to
determine the distance from the Sun to the Galactic
center [11–13]. They also serve to study the Galactic
disk warp [14–16] and the features in the distribution
of various chemical elements in the Galactic disk [17–
19].

New data on more than 2200 classical Cepheids
were presented in the catalog by Skowron et al. [20].
Distances to these stars were determined from the
period–luminosity relation with errors of 5–10
Mr\'oz et al. [21] provided more than 800 Cepheids
from these catalog with the line-of-sight velocities
from the Gaia DR2 catalog [22] and determined the
Galactic rotation parameters from these data. In the
paper by Bobylev et al. [22], it was concluded that the
distances to Cepheids obtained by Skowron et al. [20]
should be made longer by 9\%, while the kinematic
parameters were estimated without correcting the distance
scale [20].

When analyzing the kinematics of stars, it is
important to have in hand high-accuracy values for
their trigonometric parallaxes and proper motions. At
present, there is a version of the Gaia EDR3 catalog
[23] containing the trigonometric parallax values for
1.5 billion stars specified more accurately by approximately
30\%, as compared to those in the previous
Gaia DR2 version; and the accuracy of the proper
motions of these stars were improved roughly twice.
The trigonometric parallaxes of about one third of
stars from the Gaia EDR3 catalog were measured with
errors smaller than 0.2 milliarcsecond (mas). The
proper motions of about half of stars from this catalog
were measured with an error less than 10\%.

We note that the distances to stars calculated
through trigonometric parallaxes are reliable only if
the errors in parallaxes are small (less than 10
errors in parallaxes are high, the techniques, which are
similar to the method developed by Lutz and Kelker
[24], are used. With these techniques, the distances to
about 1.47 billion stars from the Gaia EDR3 catalog,
which had been calculated through trigonometric parallaxes, 
were improved by Bailer-Jones et al. [25]. In
this study, we use the distances to Cepheids from the
paper [25] as one of the sources of the distances.

The purpose of this study is to compare the distances
to Cepheids, which were obtained from the
period–luminosity relations in a paper [20], with the
distances based on trigonometric parallaxes from the
Gaia catalog. It would be also of interest to redefine
the Galactic rotation parameters and the parameters
of a spiral density wave with the use of a more accurate
Cepheid distance scale.

 \section*{METHOD}\label{method}
From observations, we have three components of
the velocity of a star: the line-of-sight velocity $V_r$ and
two projections of the tangential velocity $V_l=4.74r\mu_l\cos b$
and $V_b=4.74r\mu_b$ directed along the Galactic longitude $l$ and latitude $b$, respectively.
Each of these three velocities is expressed in kilometers
per second; the coefficient 4.74 is a dimension
ratio, and $r$ is the heliocentric distance of a star
expressed in kiloparsecs. The proper motion components $\mu_l\cos b$
and $\mu_b$ are expressed in milliarcsecons
per year (mas/yr). The components $V_r, V_l,$ and $V_b$ are
used to calculate the velocities $U,V,$ and $W$ directed
along the axes of the Galactic rectangular coordinate
system:
 
 \begin{equation}
 \begin{array}{lll}
 U=V_r\cos l\cos b-V_l\sin l-V_b\cos l\sin b,\\
 V=V_r\sin l\cos b+V_l\cos l-V_b\sin l\sin b,\\
 W=V_r\sin b                +V_b\cos b,
 \label{UVW}
 \end{array}
 \end{equation}
where the velocities $U, V,$ and $W$ are oriented from the
Sun to the Galactic center, along the Galactic rotation
direction, and to the Northern Galactic pole, respectively.
Two velocities -- $V_R$ directed radially from the
Galactic center and the orthogonal one $V_{circ}$ directed
along Galactic rotation -- can be determined from the
following relationships 
 \begin{equation}
 \begin{array}{lll}
  V_{circ}= U\sin \theta+(V_0+V)\cos \theta, \\
       V_R=-U\cos \theta+(V_0+V)\sin \theta,
 \label{VRVT}
 \end{array}
 \end{equation}
where the position angle $\theta$ satisfies the relationship
$\tan\theta=y/(R_0-x)$, and $x,y,z$ are the rectangular
heliocentric coordinates of a star (the velocities $U,V,W$)
are directed along the corresponding axes $x,y,z$); and $V_0$ is the linear velocity of Galactic rotation at the solar distance $R_0$. 

To determine the parameters of the Galactic rotation
curve, we use the equations derived from Bottlinger’s
formulas, in which the angular velocity $\Omega$ was
expanded in a series to terms of the second order of
smallness in $r/R_0$:
\begin{equation} \begin{array}{lll}
 V_r=-U_\odot\cos b\cos l-V_\odot\cos b\sin l\\
 -W_\odot\sin b+R_0(R-R_0)\sin l\cos b\Omega^\prime_0\\
 +0.5R_0(R-R_0)^2\sin l\cos b\Omega^{\prime\prime}_0,
 \label{EQ-1}
 \end{array} \end{equation}
\begin{equation} \begin{array}{lll}
 V_l= U_\odot\sin l-V_\odot\cos l-r\Omega_0\cos b\\
 +(R-R_0)(R_0\cos l-r\cos b)\Omega^\prime_0\\
 +0.5(R-R_0)^2(R_0\cos l-r\cos b)\Omega^{\prime\prime}_0,
 \label{EQ-2}
 \end{array}\end{equation}
\begin{equation} \begin{array}{lll}
 V_b=U_\odot\cos l\sin b + V_\odot\sin l \sin b\\
 -W_\odot\cos b-R_0(R-R_0)\sin l\sin b\Omega^\prime_0\\
    -0.5R_0(R-R_0)^2\sin l\sin b\Omega^{\prime\prime}_0,
 \label{EQ-3}
 \end{array} \end{equation}
where $R$ is the distance from the star to the Galactic
rotation axis and $R^2=r^2\cos^2 b-2R_0 r\cos b\cos l+R^2_0.$
The velocities $(U,V,W)_\odot$ are the mean group
velocity of a sample; since they are representative of
the peculiar motion of the Sun, they are taken with
opposite signs. $\Omega_0$ is the angular velocity of Galactic
rotation at the solar distance $R_0$; and 
the parameters $\Omega^{\prime}_0$ and $\Omega^{\prime\prime}_0$ 
are the corresponding derivatives of the
angular velocity $V_0=R_0\Omega_0$. In this paper, the value of $R_0$
is assumed to be $8.1\pm0.1$ kpc according to a review
[26], where it was derived as a weighted average value
from a large number of up-to-date individual estimates. 
 
By solving conditional equations of form (\ref{EQ-1})–(\ref{EQ-3})
with the least squares method (LSM), we can find the
following six unknown quantities: $(U,V,W)_\odot,$ $\Omega_0$, $\Omega^{\prime}_0$, 
and $\Omega^{\prime\prime}_0$. When solving only one of the conditional
equations of form (\ref{EQ-1}) with the LSM, we may find only
five unknown quantities: $(U,V,W)_\odot,$ $\Omega^{\prime}_0$, and $\Omega^{\prime\prime}_0$.

The impact of a spiral density wave on the radial $V_R$
and residual tangential velocities $\Delta V_{circ}$ is periodic,
and its amplitude is about 10 km/s. According to the
linear theory of density waves by Lin and Shu [27], this
influence is described by the relationships of the following
form:
 \begin{equation}
 \begin{array}{lll}
       V_R =-f_R \cos \chi,\\
 \Delta V_{circ}= f_\theta \sin\chi,
 \label{DelVRot}
 \end{array}
 \end{equation}
where
 \begin{equation}
 \chi=m[\cot(i)\ln(R/R_0)-\theta]+\chi_\odot
 \end{equation}
is the spiral wave phase; $m$ is the number of spiral
arms; $i$ is a pitch angle of the
spiral pattern ( $i<0$ for a coiling spiral); $\chi_\odot$ is the radial
phase of the Sun in a spiral wave; while $f_R$ and $f_\theta$ are
the amplitudes of perturbations in the radial and tangential velocities, 
respectively, which are assumed to be positive.
 
To study the periodicities in the velocities $V_R$ and
$\Delta V_{circ}$, we use the spectral (periodogram) analysis.
The wavelength $\lambda$ (the distance between neighboring
segments of spiral arms counted along the radial direction)
is calculated from the relationship
\begin{equation}
 2\pi R_0/\lambda=m\cot(|i|).
 \label{a-04}
\end{equation}
Let us consider a number of the measured velocities $V_{R_n}$
(they may be radial $V_R,$ tangential $\Delta V_{circ}$, or vertical $W$
velocities), where $n=1,\dots,N$, and $N$ is the number
of objects. The purpose of the spectral analysis is
to reveal the periodicity from a data set in accordance
with the assumed model describing a spiral density
wave with the parameters $f$, $\lambda$~(or $i$), and $\chi_\odot$.

Due to accounting for the logarithmic character of
a spiral wave and the position angles of objects $\theta_n$, our
spectral analysis of the velocity perturbation series is
reduced to calculations of the square of the amplitudes
(the power spectrum) of the standard Fourier transformation
[28]:
\begin{equation}
 \bar{V}_{\lambda_k} = \frac{1} {N}\sum_{n=1}^{N} V^{'}_n(R^{'}_n)
 \exp\biggl(-j\frac {2\pi R^{'}_n}{\lambda_k}\biggr),
 \label{29}
\end{equation}
where $\bar{V}_{\lambda_k}$~is the $k$-th harmonic of the Fourier transformation
with the wavelength $\lambda_k=D/k$; $D$ is the period
of the analyzed series; and
 \begin{equation}
 \begin{array}{lll}
 R^{'}_{n}=R_0\ln(R_n/R_0),\\
 V^{'}_n(R^{'}_n)=V_n(R^{'}_n)\times\exp(jm\theta_n).
 \label{21}
 \end{array}
\end{equation}
The required wavelength $\lambda$ corresponds to a peak value 
of the power spectrum $S_{peak}$. The pitch angle of a spiral
density wave is calculated from Eq. (\ref{a-04}). The amplitude
and the phase of perturbations are obtained from fitting
the measured data by the harmonic with the wavelength
determined.

\begin{figure}[t]
{ \begin{center}
  \includegraphics[width=0.5\textwidth]{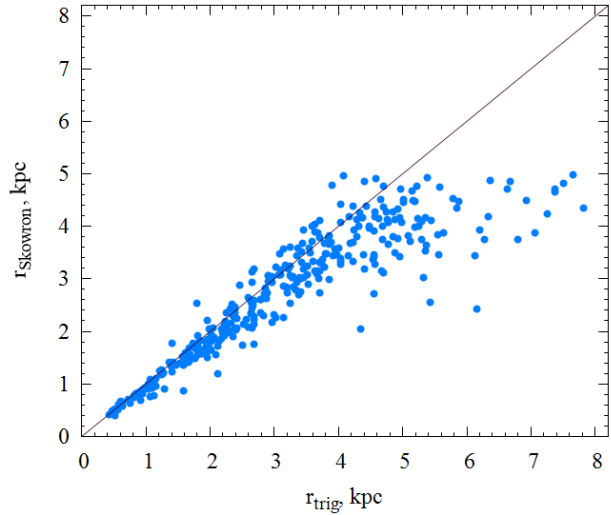}
  \caption{
The distances to Cepheids from a paper [20] versus the distances calculated through the Gaia EDR3 trigonometric parallaxes without the correction $\Delta\pi$; the diagonal line of coincidence is shown.
  }
 \label{f-dist-1}
\end{center}}
\end{figure}
 \begin{table}[t] \caption[]{\small 
Kinematic parameters found by Cepheids with the use of the distances from a paper [20].
The parameters obtained on the base of the period–luminosity relation (see [20]) are in the upper part of the table. The parameters obtained from the distances to Cepheids, which were lengthened by 10\%, are in the lower part of the table.
 }
  \begin{center}  \label{t:01}
  \small
  \begin{tabular}{|l|r|r|r|r|r|}\hline
  Parameters  & $V_r+V_l+V_b$ &  $V_r$ & $V_l$ \\\hline
  &&&\\
   $U_\odot,$ km/s & $ 8.15\pm0.63$ & $ 9.57\pm1.25$ & $8.08\pm0.94$ \\
   $V_\odot,$ km/s & $12.87\pm0.80$ & $15.53\pm1.13$ & $7.47\pm1.77$ \\
   $W_\odot,$ km/s & $ 7.25\pm0.61$ &            --- &           --- \\

   $\Omega_0,$ km/s/kpc &$ 29.03\pm0.25$ & --- & $29.17\pm0.35$ \\
  $\Omega^{'}_0,$ km/s/kpc$^{2}$
      &$-4.015\pm0.066$& $-4.216\pm0.101$&$-3.747\pm0.123$\\
 $\Omega^{''}_0,$ km/s/kpc$^{3}$
      &$ 0.639\pm0.052$& $ 0.700\pm0.106$& $0.417\pm0.087$\\
   $\sigma_0,$ km/s &       $11.60$ &       $14.12$ &       $13.30$ \\
  \hline
   $U_\odot,$ km/s & $ 9.03\pm0.65$ & $ 9.55\pm1.24$ & $8.77\pm1.02$ \\
   $V_\odot,$ km/s & $11.58\pm0.81$ & $12.78\pm1.12$ & $6.89\pm1.91$ \\
   $W_\odot,$ km/s & $ 8.01\pm0.62$ &            --- &           --- \\

   $\Omega_0,$ km/s/kpc &$ 28.87\pm0.23$ & --- & $29.25\pm0.35$ \\
  $\Omega^{'}_0,$ km/s/kpc$^{2}$
      &$-3.894\pm0.063$& $-3.893\pm0.093$&$-3.887\pm0.126$\\
 $\Omega^{''}_0,$ km/s/kpc$^{3}$
      &$ 0.602\pm0.044$& $ 0.593\pm0.088$& $0.519\pm0.081$\\
   $\sigma_0,$ km/s &       $11.89$ &       $13.99$ &       $14.48$ \\
 \hline
\end{tabular}\end{center} \end{table}

 \begin{table}[t] \caption[]{\small
Kinematic parameters determined by 308 Cepheids with the fundamental pulsation mode.
The distances from a paper [20] were lengthened by 10\%. 
 }
  \begin{center}  \label{t:02}
  \small
  \begin{tabular}{|l|r|r|r|r|r|}\hline
  Parameters  & $V_r+V_l+V_b$ &  $V_r$ & $V_l$ \\\hline

   $U_\odot,$ km/s & $ 8.77\pm0.69$ & $ 9.54\pm1.34$ & $8.34\pm1.08$ \\
   $V_\odot,$ km/s & $12.68\pm0.88$ & $13.81\pm1.23$ & $7.98\pm2.01$ \\
   $W_\odot,$ km/s & $ 8.27\pm0.79$ &            --- &           --- \\

   $\Omega_0,$ km/s/kpc &$ 28.52\pm0.25$ & --- & $28.99\pm0.37$ \\
  $\Omega^{'}_0,$ km/s/kpc$^{2}$
      &$-3.877\pm0.068$& $-3.870\pm0.099$&$-3.911\pm0.136$\\
 $\Omega^{''}_0,$ km/s/kpc$^{3}$
      &$ 0.641\pm0.047$& $ 0.619\pm0.093$& $0.575\pm0.085$\\

   $\sigma_0,$ km/s &       $11.84$ &       $14.00$ &       $14.27$ \\
 \hline
\end{tabular}\end{center} \end{table}

 \section*{DATA}\label{data}
In this study, we use the data on classical Cepheids
reported by Skowron et al. [20], who estimated the distances
and the ages of 2431 Cepheids. These Cepheids
were observed in the frames of the fourth stage of the
Optical Gravitational Lensing Experiment (OGLE)
program [29]. Skowron et al. [20] calculated the distances
to Cepheids on the base of the period–luminosity
relation from the light curves in the mid-IR
range. According to these authors, in their catalog,
random errors in determining the distances to Cepheids
are within an interval of 5-10\%. In a paper [20],
the age of Cepheids was estimated by the technique
proposed by Anderson et al. [30] with accounting for
the axial rotation of stars and their metallicity.

We used the stars from the paper by Mr\'oz et al.
[21], who provided a number of Cepheids from the list
[20] with the line-of-sight velocities and the proper
motions taken from the Gaia DR2 catalog. For the
kinematic analysis, we selected the Cepheids younger
than 120 Myr, which are located no further than 5 kpc
from the Sun, and identified them with objects from
the Gaia EDR3 catalog and the list by Bailer-Jones
et al. [25].

In a paper [25], for approximately 1.47 billion stars
from the Gaia EDR3 catalog, the distances were more
accurately calculated through trigonometric parallaxes.
The systematic correction to the Gaia EDR3
parallaxes, the value of which, $\Delta\pi=-0.017$~mas, had
been determined by Lindegren et al. [31], was considered.
The authors used two techniques, purely geometric
(geom) and photogeometric (phgeom) ones, to
calculate two versions of the distances. According to
their opinion, the photogeometric technique results in
more precise distances. We note that the both methods
are model-dependent, since they use the referencing
to the Galaxy model. As the authors themselves
notice, these distances are useful when the errors in
determining the trigonometric parallaxes of stars from
the Gaia EDR3 catalog are high.

Thus, we have 363 Cepheids. For them, there are
distances obtained from the period--luminosity relation,
as well as distances based on trigonometric parallaxes
from the Gaia EDR3 catalog. At the same
time, a number of distance scales can be used. First,
these are the distances calculated directly from the
Gaia EDR3 parallaxes. Second, these are two kinds of
the distances to Cepheids taken from a paper [25].

\begin{figure}[t]
{ \begin{center}
  \includegraphics[width=0.85\textwidth]{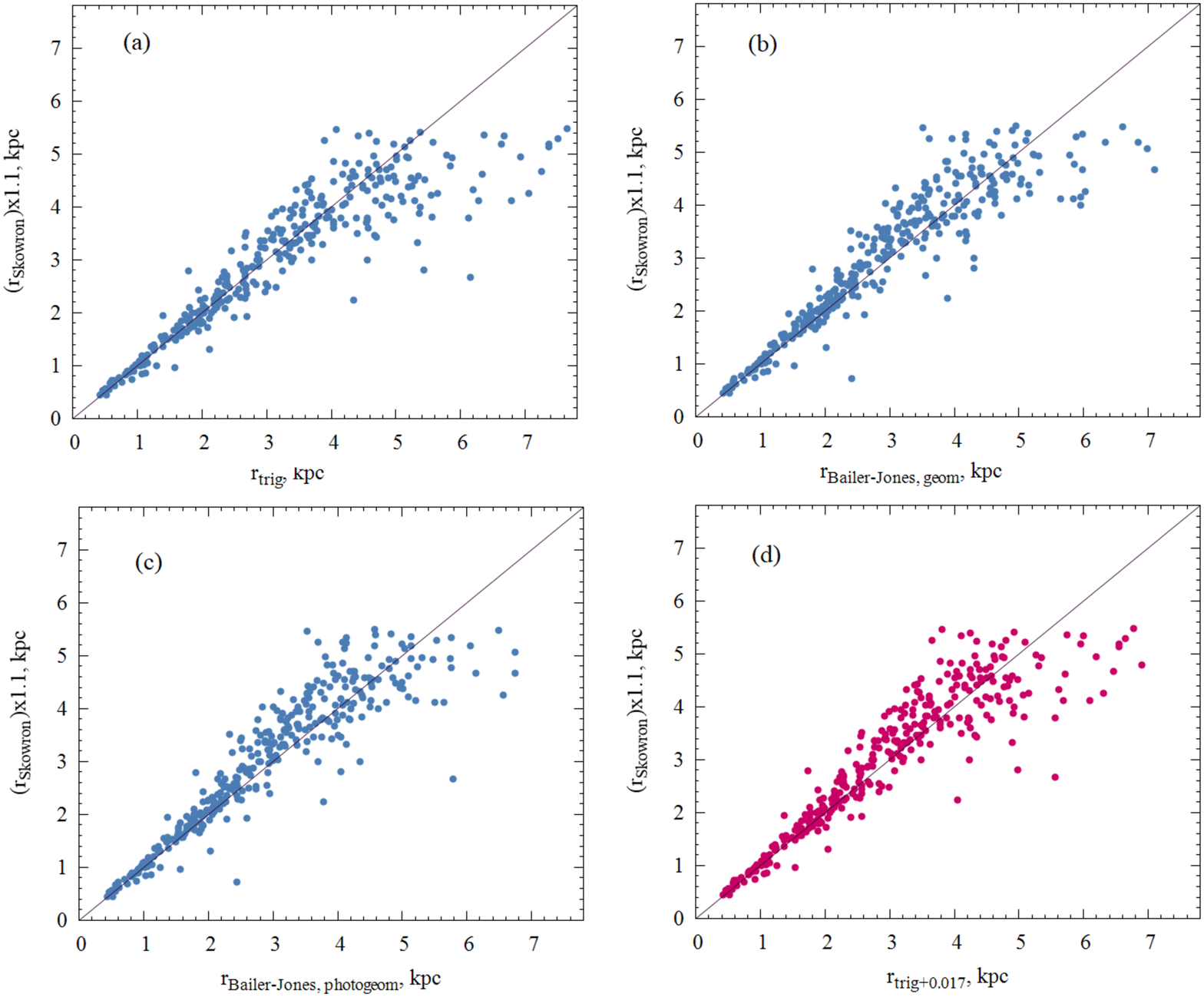}
  \caption{
The distances to Cepheids from a paper [20] lengthened by 10\% versus the distances calculated through the Gaia EDR3 trigonometric parallaxes (a), the geometric distance from a paper [25] (b), the photo-geometric distance from a paper [25], and the distances calculated through the Gaia EDR3 trigonometric parallaxes with accounting for the correction $\Delta\pi=-0.017$~mas (d). In each of the panels, the diagonal line of coincidence is shown.
  }
 \label{f-dist-2}
\end{center}}
\end{figure}
\begin{figure}[t]
{ \begin{center}
  \includegraphics[width=0.5\textwidth]{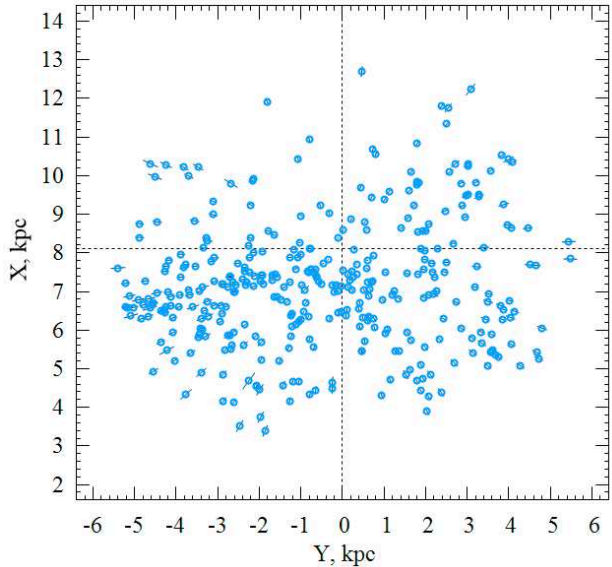}
  \caption{
The distribution of the considered sample of Cepheids in projection onto the Galactic $XY$ plane.  
  }
 \label{f-XY}
\end{center}}
\end{figure}
\begin{figure}[t]
{ \begin{center}
  \includegraphics[width=0.95\textwidth]{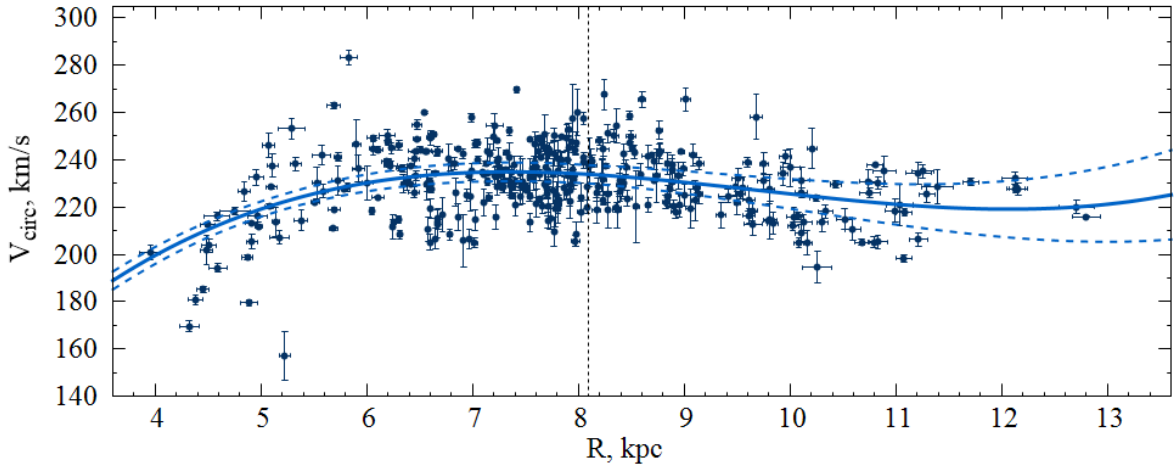}
  \caption{
The circular rotation velocities of Cepheids $V_{circ}$ versus the distances $R$. The Galactic rotation curve is shown with the confidence region bordered at the 1? level, and the position of the Sun is marked by a vertical line.
  }
 \label{f4-rotation}
\end{center}}
\end{figure}
\begin{figure}[t]
{ \begin{center}
  \includegraphics[width=0.85\textwidth]{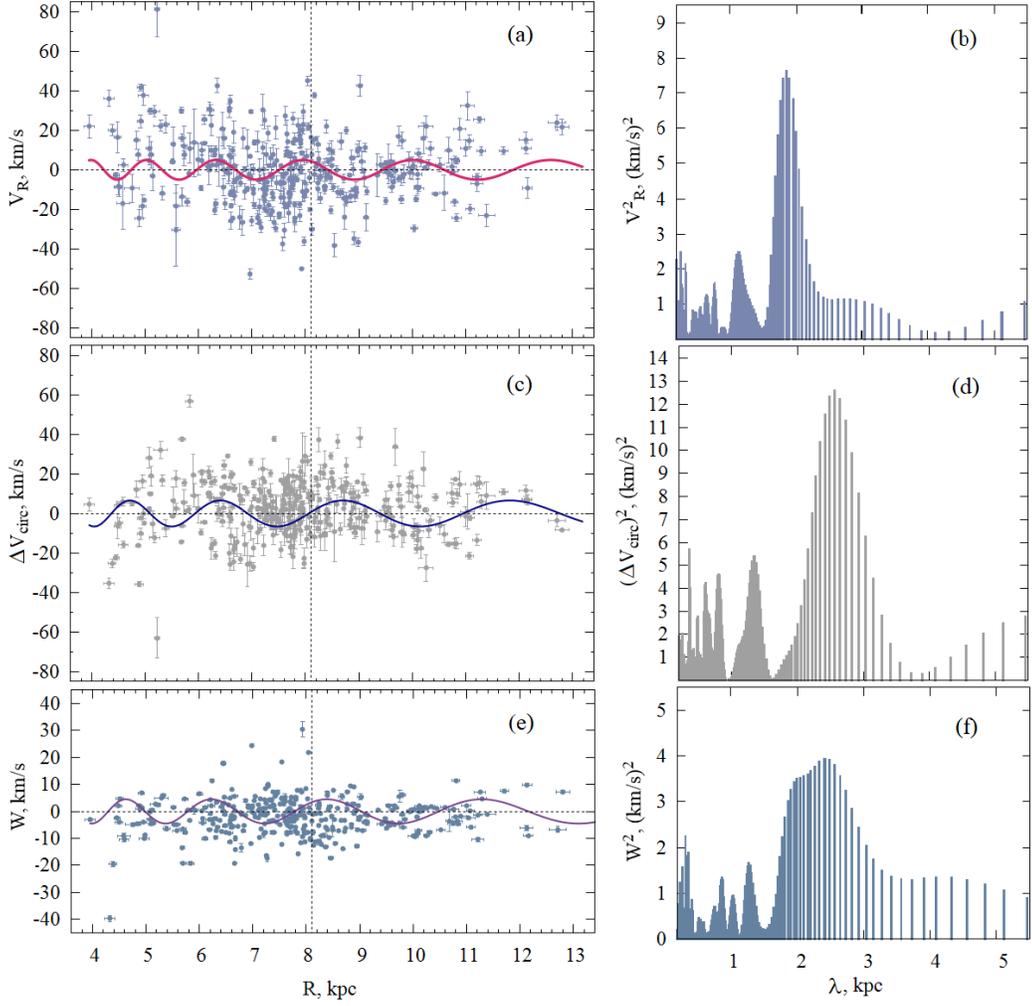}
  \caption{
The radial velocities $V_R$ versus the distances $R$ for the considered sample of Cepheids (a) and the power spectrum of this sample (b); the residual rotation velocities of Cepheids $\Delta V_{circ}$ (c) and their power spectrum (d); and the vertical velocities of Cepheids $W$ (e) and their power spectrum (f).
  }
 \label{f5-Spectr}
\end{center}}
\end{figure}
\begin{figure}[t]
{ \begin{center}
  \includegraphics[width=0.85\textwidth]{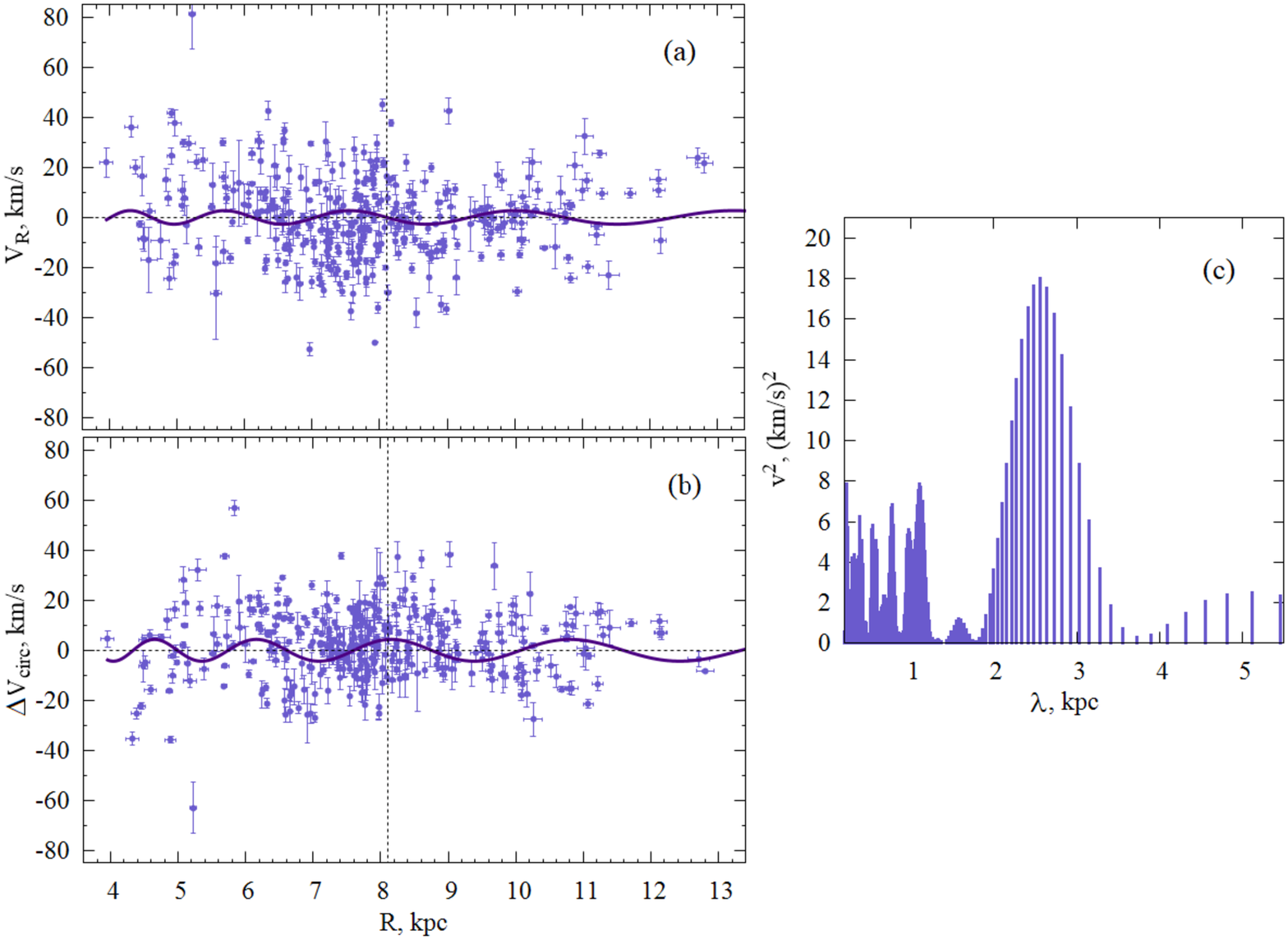}
  \caption{
The radial velocities $V_R$~(a) and the residual rotation velocities of Cepheids $\Delta V_{circ}$ (b) versus the distances $R$; the power spectrum obtained by using the joint approach (c).
  }
 \label{f6}
\end{center}}
\end{figure}

 \section*{RESULTS AND DISCUSSION}
We solved conditional equations of form (\ref{EQ-1})--(\ref{EQ-3}) in
three ways. First, three velocities $V_r+V_l+V_b$ were
jointly used. Second, only the line-of-sight velocities $V_r$
of Cepheids were used. Third, only the velocities $V_l$
were used.

In the upper part of Table~\ref{t:01}, we show the kinematic
parameters obtained by using the distances to Cepheids,
which were calculated on the base of the period–
luminosity relation in a paper [20].

According to papers [32, 33], the ratio of the first-order
derivative of the Galactic rotation angular velocity
obtained only from the proper motions to that
obtained only from the line-of-sight velocities yields
the coefficient of the distance scale $p=(\Omega^{'}_0)_{V_l}/(\Omega^{'}_0)_{V_r}$ .
This coefficient serves as a correction factor of the
type $p=r/r_{true}$, where $r$ and $r_{true}$ stand for the used
and actual distances, respectively; hence, $r_{true}=r/p$.
The error in the coefficient $p$ was calculated from the
relationship $\sigma^2_p=(\sigma_{\Omega'_{0V_l}}/\Omega'_{0V_r})^2+
     (\Omega'_{0V_l}\cdot\sigma_{\Omega'_{0V_r}}/\Omega'^2_{0V_r})^2$.
According to the data from Table~\ref{t:01}, we find $p=0.89\pm0.04$.
This means that the distances to
Cepheids calculated on the base of the period–luminosity
relation in a paper [20] should be lengthened
approximately by 10\%.

In the lower part of Table~\ref{t:01}, we present the kinematic
parameters obtained by using the distances
already lengthened by 10\%. The results of comparing
different scales of distances to Cepheids are shown in
Figs.~\ref{f-dist-1} and \ref{f-dist-2}. In these diagrams, the distances from a
paper [20] are shown in dependence on those calculated
from the Gaia EDR3 trigonometric parallaxes
($r=1/\pi$), the geometric and photo-geometric distances
from a paper [25], as well as the distances calculated
from the Gaia EDR3 trigonometric parallaxes
with accounting for the correction $\Delta\pi=-0.017$~mas
($r=1/(\pi+0.017)$).

It is seen from Fig.~\ref{f-dist-1} that the scale of distances to
Cepheids obtained through the Gaia EDR3 trigonometric
parallaxes is substantially longer than the scale
by Skowron et al. [20]. Moreover, even for small distances,
the agreement between these scales is not quite
good.

From a comparison of the distances to Cepheids
shown in Fig.~\ref{f-dist-2}, we may conclude the following: (1) in
general, the distances from a paper [20] and the scales
based on the Gaia EDR3 trigonometric parallaxes well
agree; (2) the worst agreement is observed in panel (a)
of this figure, and this is most evident when the distances
from the Sun exceed 4 kpc; (3) the distances
shown in panels (b) and (c) almost completely agree;
(4) after correcting the distance scales shown in
Fig.~\ref{f-dist-2}d, they become well consistent with each other;
this circumstance is especially important for the scale
by Skowron et al. [20], with which the kinematic
parameters are estimated in the present study.

Figure ~\ref{f-XY} shows the distribution of Cepheids
younger than 120 Myr in projection onto the Galactic
plane $XY$. In this coordinate system, the axis $X$ is
directed from the Galactic center to the Sun, while the
direction of the axis $Y$ coincides with the Galactic
rotation direction. Hereinafter, the distances to
Cepheids from the list [20] were made longer by 10\%.

Figure~\ref{f4-rotation} shows the circular rotation velocities of
Cepheids $V_{circ}$ in dependence on the distance $R$. We
present the Galactic rotation curve, which is constructed
with the parameters given in the lower part of
Table~\ref{t:01} (second column). This Galactic rotation curve
was used to calculate the residual circular rotation
velocities of Cepheids $\Delta V_{circ}$.

The Galactic rotation parameters determined in
the present study well agree with their estimates
obtained by using the other Galactic sources. For
example, from the data on 130 galactic masers with the
measured trigonometric parallaxes, the authors of a
paper [33] found the velocity components for the Sun
$(U_\odot,V_\odot)=(11.40,17.23)\pm(1.33,1.09)$~km/s and
the following values for the Galactic rotation curve
parameters: $\Omega_0=28.93\pm0.53$~km/s/kpc,
$\Omega^{'}_0=-3.96\pm0.07$~km/s/kpc$^{2}$,  $\Omega^{''}_0=0.87\pm0.03$~km/s/kpc$^{3}$,
and  $V_0=243\pm10$~km/s for a value of $R_0=8.40\pm0.12$~kpc obtained.

We note that the catalog [20] contains Cepheids
exhibiting pulsations in both the fundamental mode
and the first overtone. For each of these pulsation
modes, slightly differing calibrations were used, which
may influence the distance scales.

Among 363 Cepheids selected for this analysis,
those with the fundamental pulsation mode strongly
dominate: their number is 308, while there are only
55 Cepheids with pulsations on the first overtone. In
Table~\ref{t:02}, we present the kinematic parameters obtained
by 308 Cepheids with the fundamental pulsation
mode. The values of the kinematic parameters from
Table~\ref{t:02} should be compared to those in the lower part
of Table~\ref{t:01}. We note that this comparison reveals no
significant differences between these parameters. The
most important fact is that, according to the data in
Table~\ref{t:02}, the value of the scaling factor $p$ is
$1.01\pm0.04$. This suggests that these Cepheids are completely 
consistent with the correction of the scale by Skowron
et al. [20] performed here. Thus, there are no differences
between the results for the most homogeneous
(in terms of a pulsation mode) sample of Cepheids and
the mixed sample.

For the velocities $V_R,$ $\Delta V_{circ}$, and $W$, a spectral
analysis was conducted. The results --- the velocities $V_R,$ $\Delta V_{circ}$, and $W$ 
versus the distance and the corresponding
power spectra --- are presented in Fig.~\ref{f5-Spectr}.
From the velocities $V_R$ and $\Delta V_{circ}$, the following estimates
were obtained: $f_R=5.5\pm2.0$~km/s and
$f_\theta=7.1\pm2.0$~km/s; $\lambda_R=1.9\pm0.5$~kpc ($i_R=-8.3\pm2.5^\circ$ for $m=4$)
and  $\lambda_\theta=2.6\pm0.5$~kpc ($i_\theta=-11.4\pm2.8^\circ$ for $m=4$);
and $(\chi_\odot)_R=-208\pm16^\circ$ and
 $(\chi_\odot)_\theta=-185\pm18^\circ$.
The vertical velocities (Figs. ~\ref{f5-Spectr}(e) and 5f)
also exhibit periodicity, though with a smaller amplitude:
$f_W=3.9\pm2.0$~km/s, $\lambda_W=2.4\pm0.5$~kpc
($i_W=-10.7\pm2.7^\circ$ for $m=4$), and $(\chi_\odot)_W=-138\pm18^\circ$. 
It should be noted that there is a noticeable diversity
between the values of the wavelength $\lambda$ determined
by the velocities $V_R$ and $\Delta V_{circ}$.

We also performed a joint spectral analysis of the
velocities $V_R$ and $\Delta V_{circ}$. Here, for both kinds of velocity,
we assume that the wavelength and the phase of the
Sun in a density wave take the same values. The results
are shown in Fig.~\ref{f6}. This approach yielded the following
estimates: $f_R=2.7\pm1.7$~km/s,
$f_\theta=4.4\pm1.7$~km/s, and $\lambda_R=2.6\pm0.5$~kpc.

Our preference is that the separate solutions yield
more reliable results; this especially concerns the
results of the separate solution in the radial velocities.
Thus, for the joint solution (Fig.~\ref{f6}), the significance of
a peak in the power spectrum is 0.986. At the same
time, the analysis performed separately in the tangential
velocity (Fig.~\ref{f5-Spectr}d) and the radial one (Fig.~\ref{f5-Spectr}b)
yields the values 0.985 and 0.990, respectively, for the
significance of a peak in the power spectrum. This is a
formal point of view.

On the other hand, as is seen from Figs.~\ref{f5-Spectr} and \ref{f6}, the
difference in the phase of the Sun in a wave is strong.
The wave, which is most close to the expected one, is
a wave in the radial velocities found in the separate
analysis (Fig.~\ref{f5-Spectr}a). Indeed, from Fig.~\ref{f-XY} we can see that
Cepheids are strongly concentrated toward a segment
of the Carina–Sagittarius spiral arm at a distance of
$R\sim7$~kpc. According to the model by Lin and Shu
[27], the velocity perturbation $f_R$ in the center of the
spiral arm (at $R\sim7$~kpc) should be directed to the
Galactic center (i.e., it should take negative values in
the diagram), which can be observed in Fig.~\ref{f5-Spectr}a rather
than Fig.~\ref{f6}a. We note that the radial velocities are most
important for this problem, since they are independent
of the accounting for the rotation curve.

The parameters found in this study may be compared
to the estimates obtained in a paper [13] from a
sample of Cepheids younger than 90 Myr (their distances
are also from the paper by Skowron et al. [20],
but the proper motions are from the Gaia DR2 catalog): $f_R=12.0\pm2.6$~km/s,
$f_\theta=8.9\pm2.5$~ km/s,
$\lambda_R=2.5\pm0.3$~kpc ($i_R=-10.8\pm3.1^\circ$ for $m=4$), and
$\lambda_\theta=2.7\pm0.5$~kpc ($i_\theta=-11.8\pm3.1^\circ$ for $m=4$). This
comparison reveals a complete agreement between the
estimates of $f_\theta,\lambda_\theta$, and $i_\theta$ in the tangential direction
and a less agreement between the estimates in the
radial direction. The difference is caused by the samples’
arrangement. In our case, the sample is limited
by the distance and dominated by Cepheids from the
Carina-Sagittarius spiral arm segment (Fig.~\ref{f-XY}).

It is also worth mentioning the paper by Loktin and
Popova [34], where the authors analyzed the kinematics
of $\sim$1000 open star clusters (OSCs) of different ages with the proper motions of stars from the
Gaia DR2 catalog, which yielded the following estimates:
$f_R=4.6\pm0.7$~km/s and $f_\theta=1.1\pm0.4$~km/s.

In a paper [35], a spectral analysis was applied to
the spatial velocities of 233 young OSCs with the
proper motions and parallaxes from the Gaia EDR3
catalog. It was shown that the values for the wavelength
and the velocity perturbation determined independently
by each of the velocity kinds generally
agree, and the following estimates were obtained:
$\lambda_R=3.3\pm0.5$~kpc and $\lambda_\theta=2.6\pm0.6$~kpc, where
$i_R=-14.5\pm2.1^\circ$ and $i_\theta=-11.4\pm2.6^\circ$ for
$m=4$ and $R_0=8.1\pm0.1$~kpc assumed. The amplitudes of
the radial and tangential velocity perturbations
amounted to $f_R=9.1\pm0.8$~km/s and
$f_\theta=4.6\pm1.2$~km/s, respectively.

\section*{CONCLUSIONS}
We compared the distances to Cepheids obtained
from the period–luminosity relation to the distances
based on the trigonometric parallaxes of the Gaia catalog.
For this purpose, we used the sample of young
Cepheids from the paper by Skowron et al. [20]. It
contains 363 Cepheids younger than 120 Myr, which
are located no further than 5 kpc from the Sun. It has
been shown that the scale by Skowron et al. [20]
should be lengthened by approximately 10\%.

It has been also shown that the corrected scale by
Skowron et al. [20] well agrees with the scale that is
based on the trigonometric parallaxes from the Gaia
EDR3 catalog and contains the systematic correction
$\Delta\pi$. The corrected scale by Skowron et al. [20] is also
well consistent with the scales from the paper by
Bailer-Jones et al. [25].

By correcting the scale of Skowron et al. [20], we
found new estimates for the group velocity of the Sun
and the Galactic rotation parameters. To estimate the
parameters of the Galactic spiral density wave, a spectral
analysis of the velocities $V_R$ and $\Delta V_{circ}$ was performed.

When jointly solving the kinematic equations, we
obtained the following estimates: $(U_\odot,V_\odot,W_\odot,)=(9.03,11.58,8.01)\pm(0.65,0.81,0.62)$~km/s,
$\Omega_0=28.87\pm0.23$~km/s/kpc, $\Omega^{'}_0=-3.894\pm0.063$~km/s/kpc$^2$,
and $\Omega^{''}_0=0.602\pm0.044$~km/s/kpc$^3$, where an error in
the unit of weight was $\sigma_0=11.89$~km/s and $V_0=233.9\pm3.4$~
km/s (for the assumed distance $R_0=8.1\pm0.1$~kpc). Here, we used the distances to Cepheids from
a paper [20] multiplied by a coefficient of 1.1. In the
considered sample, Cepheids with the fundamental
pulsation mode amount to 85\%. It has been shown
that there are no significant differences between the
estimates of the kinematic parameters found with the
most homogeneous (in terms of a pulsation mode)
sample and the mixed sample of Cepheids.

A separate spectral analysis of the radial $V_R,$
residual tangential $\Delta V_{circ}$, and vertical $W$ velocities
yielded the following estimates: $f_R=5.5\pm2.0$~km/s,
$f_\theta=7.1\pm2.0$~km/s, $f_W=3.9\pm2.0$~km/s, 
 $\lambda_R=1.9\pm0.5$~kpc ($i_R=-8.3\pm2.5^\circ$ for $m=4$),
 $\lambda_\theta=2.6\pm0.5$~kpc ($i_\theta=-11.4\pm2.8^\circ$ for $m=4$),
 $\lambda_W=2.4\pm0.5$~kpc ($i_W=-10.7\pm2.7^\circ$ for $m=4$),
 $(\chi_\odot)_R=-208\pm16^\circ$,
 $(\chi_\odot)_\theta=-185\pm18^\circ$ and
 $(\chi_\odot)_W=-138\pm18^\circ$.
This analysis was also performed with the distances to
Cepheids from a paper [20] multiplied by a coefficient
of 1.1.

\section*{ACKNOWLEDGMENTS}
The authors are grateful to the referee for useful comments,
which resulted in a much improved manuscript.

 \bigskip\medskip{\Large\bf REFERENCES}\medskip {\small
 \begin{enumerate}

 \item
H.S. Leavitt, Ann. Harvard College Observ. {\bf 60}, 87 (1908).

 \item
H.S. Leavitt and E.C. Pickering, Harvard College Observ. Circ. {\bf 173}, 1 (1912).

 \item
The HIPPARCOS and Tycho Catalogues, ESA SP--1200 (1997).

 \item
M. Feast and P. Whitelock, Mon. Not. R. Astron. Soc. {\bf 291}, 683 (1997).

 \item
A.M. Mel'nik, P. Rautiainen, L.N. Berdnikov, A.K. Dambis, and A.S. Rastorguev,
 AN {\bf 336}, 70 (2015).

 \item
A.M. Mel'nik, A.K. Dambis and A.S. Rastorguev, Astron. Lett. {\bf 25}, 518 (1999).

 \item
V.V. Bobylev and A.T. Bajkova,  Astron. Lett. {\bf 38}, 638 (2012). 

 \item
A.K. Dambis, L.N. Berdnikov, Yu.N. Efremov, et al., Astron. Lett. {\bf 41}, 489 (2015).

 \item
I.I. Nikiforov and A.V. Veselova, Astron. Lett. {\bf 44}, 81 (2018).

 \item
A.V. Veselova and I.I. Nikiforov, Research Astron. Astrophys. {\bf 20}, 209 (2020).

 \item
X. Chen, S. Wang, L. Deng and R. de Grijs, Astrophys. J. {\bf 859}, 137 (2018).

 \item
D. Kawata, J. Bovy, N. Matsunaga and J. Baba, Mon. Not. R. Astron. Soc. {\bf 482}, 40 (2019).

 \item
V.V. Bobylev, A.T. Bajkova, A.S. Rastorguev, and M.V. Zabolotskikh,
 Mon. Not. R. Astron. Soc. {\bf 502}, 4377 (2021).

 \item
J.D. Fernie, Astron. J. {\bf 73}, 995 (1968).

 \item
L.N. Berdnikov, Astron. Lett. {\bf 13}, 45 (1987).

 \item
V.V. Bobylev, Astron. Lett. {\bf 39}, 753 (2013).

 \item
S.M. Andrievsky, J.R.D. L\'epine, S.A. Korotin, R.E. Luck, V.V. Kovtyukh, and W.J. Maciel, Mon. Not. R. Astron. Soc. {\bf 428}, 3252 (2013).

 \item
V.A. Marsakov, V.V. Koval’, V.V. Kovtyukh, and T.V. Mishenina,
 Astron. Astrophys. Tr. {\bf 28}, 367 (2014).

 \item
V.V. Kovtyukh, S.M. Andrievsky, R.P. Martin, et al.,
Mon. Not. R. Astron. Soc. {\bf 489}, 2254 (2019).

 \item
D.M. Skowron, J. Skowron, P. Mr\'oz, et al., Science {\bf 365}, 478 (2019).

 \item
P. Mr\'oz, A. Udalski, D.M. Skowron, et al., Astrophys. J. {\bf 870}, L10 (2019).

 \item
Gaia Collaboration, A.G.A. Brown, A. Vallenari, T. Prusti, et al.,
 Astron. Astrophys. {\bf 616}, 1 (2018). 

 \item
Gaia Collaboration, A.G.A. Brown, A. Vallenari, T. Prusti, et al.,
 Astron. Astrophys. {\bf 649}, 1 (2021). 

 \item
T.E. Lutz and D.H. Kelker, Pub. Astron. Soc. Pacific {\bf 85}, 573 (1973).

 \item
C.A.L. Bailer-Jones, J. Rybizki, M. Fouesneau, M. Demleitner, and R. Andrae, Astron. J. {\bf 161}, 147 (2021).

 \item
V.V. Bobylev and A.T. Bajkova, Astron. Rep. {\bf 65}, 498 (2021). 

 \item
C.C. Lin and F.H. Shu, Astrophys. J. {\bf 140}, 646 (1964).

 \item
A.T. Bajkova and V.V. Bobylev, Astron. Lett. {\bf 38}, 549 (2012)]. 

 \item
A. Udalski, M.K. Szyma\'nski, and G. Szyma\'nski, Acta Astron. {\bf 65}, 1 (2015).

 \item
R.I. Anderson, H. Saio, S. Ekstr\"om, C. Georgy, and G. Meynet,
Astron. Astrophys. {\bf 591}, A8 (2016).

 \item
Gaia Collaboration, L. Lindegren, S.A. Klioner, J., Hern\'andez, J., et al.,
 Astron. Astrophys. {\bf 649}, 2 (2021). 

 \item
M.V. Zabolotskikh, A.S. Rastorguev, and A.K. Dambis, Astron. Lett. {\bf 28}, 454 (2002).

 \item
A.S. Rastorguev, M.V. Zabolotskikh, A.K. Dambis, et al., Astrophys. Bulletin {\bf 72}, 122 (2017).

 \item
A.V. Loktin and M.E. Popova, Astrophys. Bull. {\bf 74}, 270 (2019).

 \item
V.V. Bobylev and A.T. Bajkova,  Astron. Lett. {\bf 48}, 9 (2022). 

 \end{enumerate} }
 \end{document}